\documentclass[letter]{aa}

\usepackage{graphicx}
\usepackage{txfonts}
\usepackage{graphicx}    % Including figure files
\usepackage{amsmath}    % Advanced maths commands
\usepackage{amsmath} %For displaying equations
\usepackage{float} %To have a [H] option for the images
\usepackage{graphicx} %To import graphics
\usepackage{booktabs} %For nicer tables
\usepackage{hyperref} %To have hyperlinks in the bibliography
\usepackage{geometry}
\usepackage{booktabs} %For nicer tables
\usepackage{ifthen}
\usepackage{lipsum}
\usepackage{color, soul} %To have \hl command
\usepackage{makecell}
\usepackage{subcaption}
\usepackage{xspace} % To have space after \nbody
\usepackage{multirow}

\usepackage[normalem]{ulem}
\newcommand{\eq}[2][]{\begin{align}
                          #2
\end{align}}

\newcommand{\lr}[1]{\left(#1\right)}

\newcommand{\BH}{\mathrm{BH}}

\newcommand{\units}[1]{\,\mathrm{#1}}

\newcommand{\sbh}{S-BH\xspace}

\newcommand{\Gaia}{{\it Gaia}\xspace}
\newcommand{\gb}{\Gaia~BH3\xspace}

\def\subinrm#1{\sb{\rm#1}}
    {\catcode`\_=13 \global\let_=\subinrm}
\mathcode`\_="8000
\def\upsubscripts{\catcode`\_=12 } 

\upsubscripts

\newcommand{\msun}{{\rm M}_\odot}
\newcommand{\feh}{{\rm [Fe/H]}}

\renewcommand{\eqref}[1]{(equation~\ref{#1})}

\begin{document}

\title{Dynamical formation of \textit{Gaia} BH3 in the progenitor globular cluster of the ED-2 stream}
\titlerunning{Dynamical formation of \gb}
\authorrunning{Marín Pina et al.}
\author{Daniel Marín Pina
\inst{1}, Sara Rastello\inst{1}, Mark Gieles\inst{1,2},  Kyle Kremer\inst{3}, Laura Fitzgerald\inst{1}, Bruno Rando Forastier\inst{1}
}

\institute{
    Departament de F\'isica Qu\`antica i Astrof\'isica, Institut de Ci\`encies del Cosmos, Universitat de Barcelona, Mart\'i i Franqu\`es 1, E-08028 Barcelona, Spain\\
    \email{danielmarin@icc.ub.edu}
\and
ICREA, Pg. Llu\'is Companys 23, E08010 Barcelona, Spain
\and
TAPIR, California Institute of Technology, Pasadena, CA 91125, USA
}

\date{Received XX; accepted XX}

\abstract{
%Context
The star--black hole (\sbh) binary known as  \gb, discovered by the \Gaia\ Collaboration  is chemically and kinematically associated with the metal-poor ED-2 stream in the Milky Way halo. }
{
%Aims
We explore the possibility that \gb\ was assembled dynamically in the progenitor globular cluster (GC) of the ED-2 stream.
} 
{
%Methods
We used a public suite of star-by-star dynamical Monte Carlo models to identify \sbh binaries in GCs with different initial masses and (half-mass) radii. 
}
{
%Results
We show that a likely progenitor of the ED-2 stream was a relatively low-mass ($\lesssim10^5\,\msun$) GC with an initial half-mass radius of $\sim 4\,$pc. Such a GC can dynamically retain a large fraction of its BH population and dissolve on the orbit of ED-2. From the  suite of models we find that GCs produce $\sim3-30$  \sbh binaries, approximately independently of initial GC mass and inversely correlated with initial cluster radius. Scaling the results to the Milky Way GC population, we find that $\sim75\%$ of the \sbh binaries formed in GCs are ejected from their host GC, all in the early phases of evolution ($\lesssim1\,$Gyr); these are expected to no longer be close to streams. The $\sim25\%$ of \sbh binaries retained until dissolution are expected to form part of streams, such that for an initial mass of the progenitor of ED-2 of a few $10^4\,\msun$, we expect $\sim2-3$ \sbh to end up in the stream. 
GC models with  metallicities similar to \gb ($\lesssim1\%$ solar) include \sbh binaries with similar BH masses ($\gtrsim30\,\msun$), orbital periods, and eccentricities.
}
{
% Conclusions
We predict that the Galactic halo contains of order $10^5$ \sbh binaries that formed dynamically in GCs, a fraction of which may readily be detected in \textit{Gaia} DR4. The detection of these sources provides valuable tests of BH dynamics in clusters and their contribution to gravitational wave sources.
}


\maketitle
%
%_____________________________________

\section{Introduction}
\label{sec:introduction}
The  discovery of a star--black hole (\sbh) binary, known as  \gb,   by the {\it Gaia} collaboration \citep{GaiaBH3}  provides a novel glimpse into the population of dormant BHs. \gb is composed of a $33 \units{\msun}$ BH with a $0.76\units{\msun}$ star. The  star is a  metal-poor giant, with $\feh\simeq-2.6$. The binary has an eccentricity of $e\simeq0.73$, and a  wide orbit with period $P\simeq4250\units{d}$. \gb is on a halo orbit, similar to the debris of the Sequoia accretion event. 

The previously discovered \Gaia BH1 \citep{elbadry_bh1} is a binary with a Sun-like star orbiting a BH with a mass $m_\BH \simeq 10\units{\msun}$, an eccentricity $e= 0.45$, and a period of $P
\simeq186\units{d}$. \Gaia BH2 \citep{el-badry_bh2} is a binary with a red giant orbiting a $m_\BH \simeq 9\units{\msun}$ BH, with an eccentricity $e \simeq 0.5$, and a relatively long period of $P\simeq1280\units{d}$.
Their metallicities ($\feh\simeq-0.2$) and their inferred orbits in the Galactic plane point towards a formation {in a young stellar population, possibly} an open cluster \citep{2023MNRAS.526..740R,2024ApJ...965...22D,Tanikawa2024,Arcasedda2024-dragon2} or an isolated stellar binary \citep{Kotko2024}.

The association of \gb with the stellar halo calls for another formation scenario. Under the assumption that the unseen companion is a single BH,\footnote{The available data cannot rule out that the unseen object is a binary BH (BBH).} 
\citet{GaiaBH3} argue that the formation of \gb in isolation is unlikely \citep[but see][]{ElBadry2024, Iorio2024}. The star does not present strong chemical peculiarities, indicating a lack of pollution by the BH progenitor in its evolved stage. This makes a dynamical formation scenario, where the system is assembled after the BH progenitor has exploded as a supernova, more plausible. Its orbit in the Galactic halo and low metallicity suggest that \gb formed in a globular cluster (GC). 

\citet{Balbinot2024} show that \gb is both chemically and kinematically associated with the ED-2 stream \citep{2023A&A...670L...2D}. 
The progenitor of the ED-2 stream was most likely a GC, given its low velocity-dispersion \citep{2023A&A...678A.115B} and the absence of an $\feh$ spread  \citep[$<0.04\units{dex}$,][]{Balbinot2024}. These authors also find a negligible spread in Na and Al,  suggesting an initial cluster mass $\lesssim 5\times10^4\,\msun$. The fact that the cluster is dissolved also provides a mass limit: using the results of \citet{2023MNRAS.522.5340G} we find that the maximum initial mass of a cluster with BHs that can dissolve in 13 Gyr on the orbit of ED-2 is $\sim9\times10^4\units{\msun}$.

To date, only two (possibly three) detached stellar-mass BH binary candidates have been found in GCs \citep[in NGC~3201,][]{2018MNRAS.475L..15G, 2019A&A...632A...3G}.\footnote{Additional BH candidates have been found in GCs as accreting sources via their X-ray/radio emission \citep[e.g.][]{Maccarone2007, Strader2012,Miller-Jones2015}.} Only line-of-sight velocities are available for these binaries, providing a minimum BH mass of $\gtrsim5\,\msun$. \citet{Kremer2018_3201} showed that these binaries were likely formed and shaped via dynamical exchange interactions. 
Finding evidence for massive BHs ($\gtrsim30\,\msun$) in a (dissolved) GC is critical to understanding the contribution of dynamically produced BBHs to the growing sample of gravitational wave sources \citep[e.g.][]{Zwart2000, Rodriguez2016,2017MNRAS.464L..36A, Antonini2023}.

In this letter we show that \gb could have formed  dynamically in the progenitor GC of the ED-2 stream. 

\section{Methods}
\label{sec:methods}
We used the suite of GC simulations run with the \textsc{Cluster Monte Carlo} (CMC) code. CMC is a H\'{e}non-type Monte Carlo code for the long-term evolution of collisional stellar systems \cite[][and references therein]{cmc}.  The integration is done on a star-by-star basis, following gravitational interactions, the internal evolution of stars and binary stars, strong few-body interactions, and dynamical ejections \citep{Weatherford2023}. Further details on the code can be found in \cite{cmc}.

For this work we used the set of 148 CMC models presented in \cite{Kremer2020}. This catalogue  is meant to sample the relevant parameter space of local-Universe GCs varying the following initial conditions: the number of stars $N$, virial radius $r_v,$\footnote{For the initial density profile of these models $r_v \simeq 1.25r_h$, where $r_h$ is the half-mass radius.} metallicity $Z$, and Galactocentric distance $R_G$. The choice of values for the above parameters is $N\in [2, 4, 8, 16] \times 10^5$, $r_v \in [0.5, 1, 2, 4]\units{pc}$, $Z\in[0.0002, 0.002, 0.02]$, and $R_G\in[2, 8, 20]\units{kpc}$, which account for 144 models, plus four extra models with $N=3.2\times 10^6$, $R_G=20\units{kpc}$, $Z \in [0.0002, 0.02]$, and $r_v\in[1, 2]\units{pc}$. All models adopt a \citet{Kroupa2001} stellar mass function in the range $0.08-150\,\msun$, resulting in an initial mean  mass of $0.6\,\msun$.

We first looked at the raw results of all \sbh occurring in the database (see Sect.~\ref{sec:origin}). 
We also made the CMC database representative for the Milky Way GC system, by applying weights to the result of GCs with different $N$ by a factor $N^{-1}$, such that the logarithmically spaced $N$ values represent an initial GC mass function of the form $M_0^{-2}$. We then only considered the low-metallicity models ($Z=2\times10^{-4}$ and $Z=2\times10^{-3}$), with relative weights 4:1, representative of the Milky Way GC population \citep[e.g.][]{Harris1996}. We considered equal contributions of the different $r_v$ and $R_G$ models.

We only considered hard binaries \citep{Heggie1975}, defined as those whose absolute value of the binding energy is greater than the average kinetic energy of particles in the GC core. This was done to exclude loosely bound pairs that are constantly being assembled and destroyed. 

All models were evolved until 14 Gyr, or until dissolution if that occured before 14 Gyr. As clusters evolve they lose mass  and stars due to stellar evolution, internal two-body relaxation, and interactions with the Galactic tidal field. Therefore, GCs dissolve over a period that depends on their mass and $R_G$. A cluster is considered dissolved when the number of particles drops below $\sim10^4$ \citep[see Sect. 2.3 of][for the exact definition]{Kremer2020}. Of the 148 models in the catalogue, 21 of them dissolved, mostly those with low masses at small Galactocentric radii (see Table 6 in \citealt{Kremer2020} for the list of dissolved models).

\section{Origin of \gb-like binaries}
\label{sec:origin}

\begin{figure*}
    \sidecaption
    \includegraphics[width=12cm]{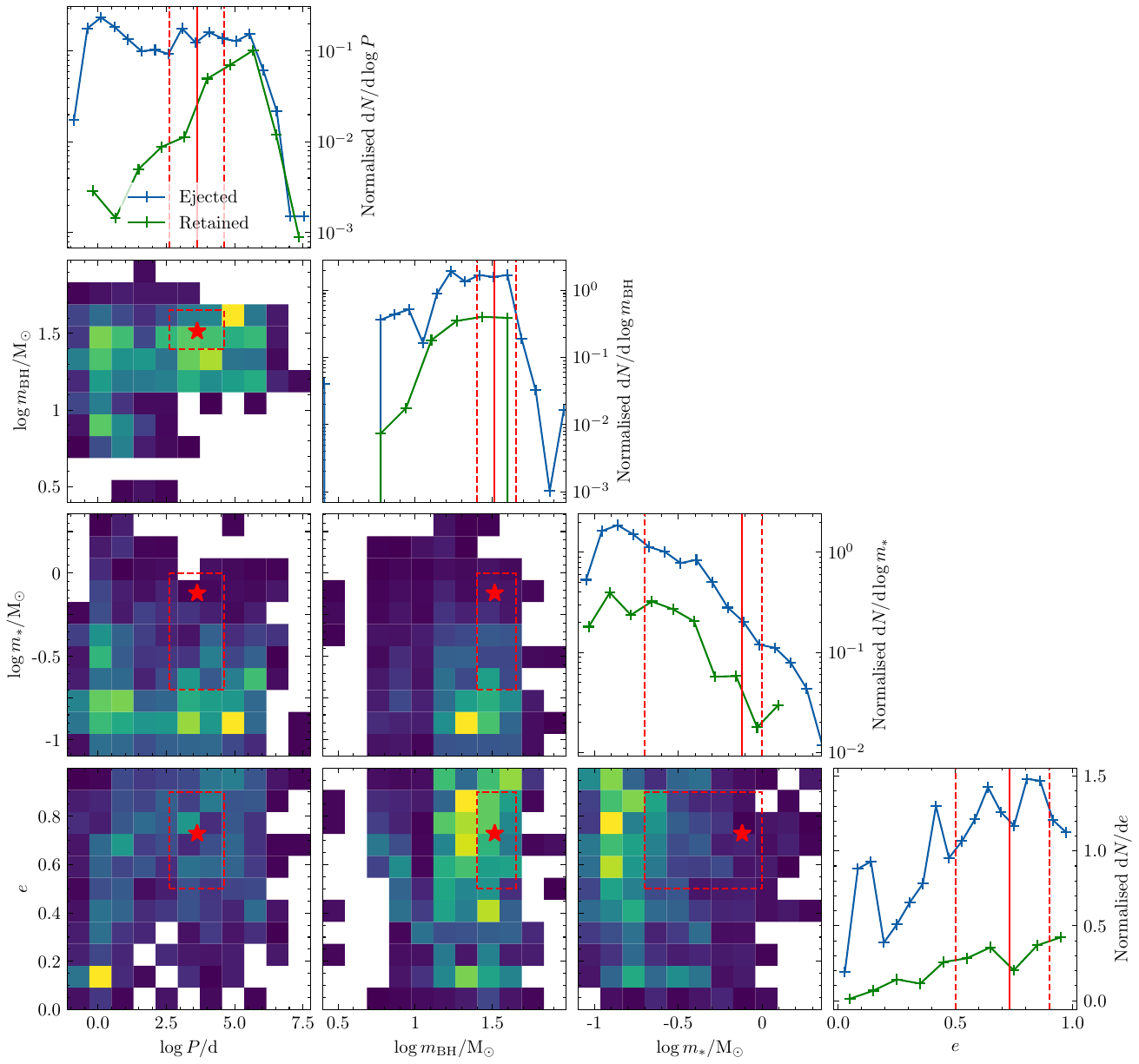}
    \caption{Probability distribution functions for the masses of the BH and star ($m_\BH$ and $m_\star$, respectively) and orbital parameters (period $P$ and eccentricity $e$) for all the detached \sbh binaries in the cluster simulations of \cite{Kremer2020}. For ejected binaries, we show their parameters at the time of ejection. The contribution of each binary to the histogram is weighted as explained in Sect.~\ref{sec:methods}. Shown in red (continuous line for 1D histograms, star symbol for 2D histograms) are the values for \gb. The  dashed red line gives  the range for the analysis of \gb-like binaries in Sect.~\ref{sec:origin}. The green lines in the 1D histogram include only the \sbh binaries that were not ejected from their parent cluster.}
    \label{fig:corner}
\end{figure*}

In this section we explore the formation of \sbh binaries in the CMC catalogue that are the most similar to \gb. From the 1837 unique \sbh binaries in the catalogue, we removed 424 accreting binaries. The remaining 1413 detached binaries are shown in Fig.~\ref{fig:corner}, where the contribution of each of the models is weighted as explained in Sect.~\ref{sec:methods}. From these binaries, 78\% were ejected and the rest remain in the cluster. We assume that the \sbh binaries that survive until 14 Gyr in surviving clusters remain in these clusters until dissolution, which is justified by the fact that most binaries form within approximately the first gigayear.

From the sample of detached binaries, we selected those most similar to \gb, which we define as those whose parameters all fall within the red dashed lines in the subplots of Fig.~\ref{fig:corner}. The red lines span the following ranges: $m_\BH\in [25, 45]\units{\msun}$, $m_\star\in [0.2, 1]\units{\msun}$, $P\in [400, 40000]\units{d}$, and $e\in [0.5, 0.9]$. These ranges are  larger than the typical uncertainties on the observationally inferred parameters of \gb,  which are of order 1-2\%,  and were   chosen to (i) cover a roughly similar fraction of the ranges of the parameters found in the models and (ii) end up with a reasonable number of \sbh\ binaries. These criteria selected 20 binaries. Next we refined this sample to match the model GC properties with what we know of the ED-2 stream, but before doing so, it is worth noting that \gb-like binaries exist and that all of them have a dynamical origin.

From the 20 resembling \sbh binaries, there were 15 ejected binaries and 5 binaries that remained in the cluster until dissolution. We refer to these five candidate \gb-like binaries as cSBH1-5. For these candidates, the masses of the components are $m_\BH=[29.9, 40.5,34.8,32.0,29.6]\units{\msun}$ and $m_\star=[0.21,0.44,0.31,0.22,0.65] \units{\msun}$. 
The semi-major axes are $a=[69,57,57,64,27] \units{AU}$ (which imply periods of $P=[38,25,27,33,9]\times 10^3\units{d}$)  and the eccentricities are $e=[0.72,0.80,0.61,0.86,0.58]$. All candidate binaries originate from cluster models with initial parameters $r_v=4\units{pc}$ and $R_{\rm G}=2\units{kpc}$. The large $r_v$  results  because clusters with low density can dynamically retain many of their BHs, which in turn leads to efficient stream formation \citep{Gieles2021,Roberts2024}. The binaries cSBH1-3 originate from the model with initial $N=2\times 10^5$, whereas cSBH4 and cSBH5 originate in the models with $N=4\times 10^5$ and $N=8\times 10^5$, respectively. If we apply the weights of the GC initial mass function (Sect.~\ref{sec:methods}), we find that three-fourths of the S-BH binaries come from the $N=2\times10^5$ models. The corresponding initial mass ($\sim10^5\,\msun$) is within a factor of $\sim2$ of the mass estimate provided by \citet{Balbinot2024}. Combined with the upper limits from the abundances and dissolution time (Sect.~\ref{sec:introduction}), we therefore conclude that the models favour an initial GC mass $\lesssim10^5\,\msun$. As expected from the high BH masses, all candidate binaries form in the lowest metallicity model, $Z=0.0002$, except for cSBH3, which has $Z=0.002$ and $m_{BH}$ grew by a stellar collision of its progenitor.

% Calculation of the maximum mass:
% ED-2 Orbit: Rperi = 6 kpc, Rapo = 20 kpc => Reff = 9.23 kpc
% Using eq 6 in GG23 with tdis = 13 Gyr, x=2/3, y = 4/3 and Mdot_ref = 45 I find Mi = 5.3e5 and assume 40% mass loss by sev I find M_0 = 9e4 Msun

Binaries cSBH1 and cSBH2 form dynamically about $100\units{Myr}$ after GC formation, corresponding to the moment of core collapse of the BH subsystem. The classic picture is that a binary forms in a three-body interaction \citep{Heggie1975, Atallah2024}. The outcome of core collapse is the production of a dynamically active BBH, likely involving between five and ten  particles \citep{Tanikawa2012}. The \sbh appears to form in an intermediate step, involving two stars and a BH, and it seems to be a byproduct of BBH formation. In the case of cSBH1, it is directly formed in a three-body interaction, whereas cSBH2 is the exchange of two three-body binaries. Developing a full understanding of the relation between the formation of the \sbh and the BBH is beyond the scope of this work. We   simply note here that it is a curiosity worth looking into in future work, and in Sect.~\ref{ssec:bbh} we discuss the implications. Binaries cSBH3-5 form in three-body interactions but at later times, around $\sim 1\units{Gyr}$. After formation, all candidate binaries undergo several strong interactions ($\sim3-18$), implying that even if formed from isolated binary evolution \citep{ElBadry2024,Iorio2024}, dynamical interactions in the ED-2 progenitor need to be included to understand their final properties.

If we consider the 15 \gb-like ejected binaries, we also find that most of them form in three-body interactions. After formation, each of the ejected binaries undergoes multiple strong interactions ($\sim1-12$)  until it is  ejected, which most do following an interaction with a BBH, which are frequent in clusters with few BHs \citep{MarinPinaGieles2024}.

\section{Implications}
\label{sec:implications}
\subsection{Formation efficiency}
\begin{figure}
    \centering
    \includegraphics[width=0.38\textwidth]{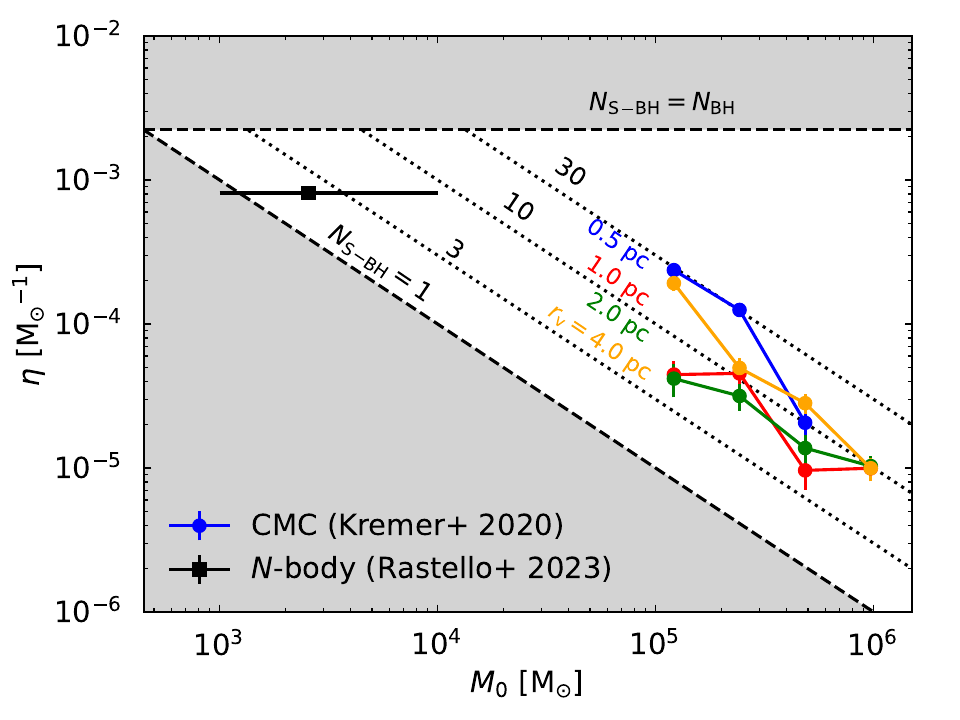}
    \caption{Formation efficiency of \sbh in all CMC models with $Z=2\times10^{-4}$ (bullets) and all \sbh formed in the $N$-body models of \citet{2023MNRAS.526..740R} for solar metallicity (square). The dashed lines show the maximum $\eta$, corresponding to all BHs in the cluster being in a  \sbh, and the minimum $\eta$, corresponding to clusters with one BH. The dotted lines correspond to constant numbers of \sbh. }
    \label{fig:eta}
\end{figure}

We now compute how many \sbh binaries are expected to be produced by the ED-2 stream. We begin by defining the formation efficiency, $\eta$, as the total number (i.e. ejected and retained) of \sbh binaries that form in a cluster per unit of $M_0$. We find a clear dependence of $\eta$ on $M_0$ and $r_v$, shown in Fig.~\ref{fig:eta}. The formation efficiency is higher in denser and less massive clusters; however, this trend does not extend to lower masses ($\lesssim 10^4\units{\msun}$). There, it levels off as there is a hard limit where the number of ejected \sbh binaries, $N_{\sbh}$, equals the initial number of BHs in the cluster, $N_\BH$ \citep[see also Fig.~1 in][]{Tanikawa2024b}. For more massive clusters, the formation of S-BH binaries is relatively inhibited by pronounced dynamical heating from their large central BH populations \citep{Kremer2018_BHstar}. In the CMC regime, the formation efficiency can be approximated by the following power law:
\eq{\eta = 1.6\times 10^{-4} \,\units{\msun^{-1}}\lr{\frac{M_0}{10^5\units{\msun}}}^{-1.2}\lr{\frac{r_{v}}{1\units{pc}}}^{-0.42}.\label{eq:eta}}
Extrapolating to the  mass estimate for the ED-2 progenitor ($5\times10^4\,\msun$, Sect.~\ref{sec:introduction}) and $r_v=2-4\,$pc, implies that the ED-2 stream produced $\sim9-13$ \sbh binaries, of which $\sim2-3$ remain in the progenitor cluster until dissolution. We conclude that the dynamical formation of \gb is a viable explanation for its origin. 

\subsection{Number of \sbh in the Galactic halo}
With the relation $\eta(M_0, r_v)$  from Eq.~(\ref{eq:eta}) we can estimate the total number of \sbh binaries produced dynamically in the complete population of Galactic GCs. 
 We write the initial GC mass function (GCMF) as $\psi_0(M_0) = AM_0^{-2}$. We then assume that all clusters lose 40\% of their mass  by stellar evolution and an amount $\Delta\simeq2\times10^5\,\msun$ by evaporation. This amount is required to obtain the turn-over mass in the GCMF at $\sim2\times10^5\,\msun$ if mass is lost at a rate that is constant in time \citep{2007ApJS..171..101J}. The evolved GCMF can then be written as   
$\psi(M) = \mu_{\rm sev}A(M+\Delta)^{-2}$, 
where $\mu_{\rm sev}\simeq0.6$.
The normalisation constant $A$ is found from the total number of Milky Way GCs (157) and integrating $\psi(M)$ from $10^2\,\msun$ (roughly the lowest mass Milky Way GC) to $10^6\,\msun$. The total number of \sbh binaries produced by all Milky Way GCs, including the dissolved ones, is then
\eq{N_{\sbh} &= \int_{10^4\,\msun}^{10^6\,\msun} \eta(M_0,r_v)M_0\psi_0(M_0) {\rm d}M_0 \simeq10^5.}
Here we weighed the results for different $r_v$ equally. We note that there are no CMC models in the range $10^4\le M_0/\msun\lesssim 10^5$, so the results depend on an  extrapolation outside the range of the CMC grid. From the models of lower mass clusters of \citet{2023MNRAS.526..740R} (Fig.~\ref{fig:eta}), this seems reasonable for $r_v\gtrsim1\,$pc. 

Because most \sbh are produced by low-mass dissolved GCs, the number density of  \sbh in the Galactic halo follows the initial density distribution of GCs in the Milky Way: $n_{GC} \propto R_G^{-4.5}$ between 1 kpc and 100 kpc \citep[e.g.][]{2001ApJ...561..751F,2023MNRAS.522.5340G}. Scaling this to a total number of $10^5$, we then  find that at the solar radius the number density of \sbh in the halo is $n_{\sbh}\simeq1\,{\rm kpc}^{-3}$, corresponding to $\sim1$ within the distance to \gb. Within a  distance of $1-3$ kpc from the Sun there are $5-130$ \sbh expected from (dissolved) GCs. We note that we predict that the ED-2 progenitor cluster has produced $\sim$ 9 -- 13 S-BH binaries, of which $\sim$ 2 -- 3 are expected to remain in the stream. The number of $n_{\sbh}\simeq1\,{\rm kpc}^{-3}$ corresponds to an average, and does not take into account the close proximity of the ED-2 stream.

\subsection{Connecting to gravitational wave sources}
\label{ssec:bbh}

As the catalogue of BBH mergers detected via gravitational waves by LIGO/Virgo/KAGRA (LVK) grows \citep{LVK2023_GWTC3}, the details of the astrophysical origin of these sources remain elusive. Dynamical formation within GCs has emerged as one possible scenario and has been studied at length using $N$-body cluster simulations, including CMC \citep[e.g.][]{Rodriguez2016}. In GCs, the dynamical formation of \sbh binaries like \gb and the three BH candidates observed in NGC~3201 is an inevitable byproduct of the same processes that lead to formation of merging BH pairs \citep[e.g.][]{Kremer2018_3201}. Thus, the detection of sources like \gb provides important constraints on the properties of BHs in clusters, their dynamics, and the role of GCs in forming GW sources.

\begin{figure}
    \centering
    \includegraphics[width=0.38\textwidth]{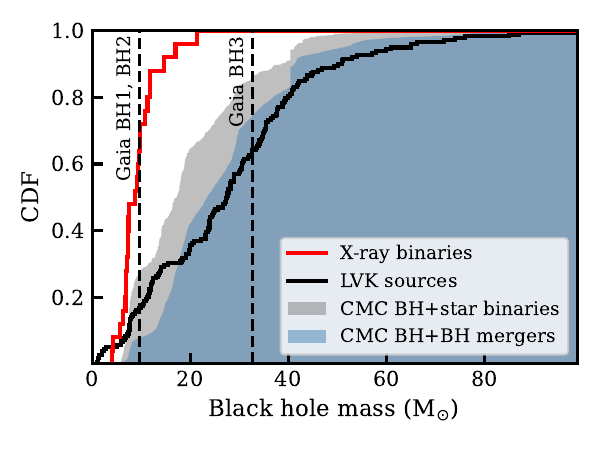}
    \caption{Cumulative distribution of measured BH masses observed as X-ray binaries \citep{Remillard2006,Corral-Santana2016} and as GW sources \citep{LVK2023_GWTC3} compared to distributions of BH masses identified in our CMC models. The vertical dashed lines show measured masses for the three \textit{Gaia} BHs.}
    \label{fig:masses}
\end{figure}

In Fig.~\ref{fig:masses} we show the cumulative distribution of BH masses from our CMC models for BHs undergoing BH mergers (grey) and BHs found in \sbh binaries (blue; including both retained and ejected binaries), again adopting the cluster weighting scheme from Sect.~\ref{sec:methods}. For reference, we also show distributions of known stellar-mass BHs with mass measurements: low- and high-mass X-ray binaries (red curve) taken from \citep{Remillard2006,Corral-Santana2016}, BH mergers (black curve) from GWTC-3 \citep{LVK2023_GWTC3}, and the three \gb binaries. The X-ray binary and GW distributions shown in Fig.~\ref{fig:masses} are raw data and do not take into account detection biases or astrophysical weighting \citep[for discussion, see][]{FishbachKalogera2022}. Figure~\ref{fig:masses}, therefore, does not suggest that the CMC models predict the observed distribution of X-ray binaries, GW sources, or \sbh binaries, merely that the dynamically active populations of BHs in the CMC models have masses similar to both the LVK sources and \gb. Future discoveries of \sbh binaries in the Milky Way, as well as detailed detectability analysis in the CMC models, can provide constraints on the  connection between these populations.

\section{Discussion}
\label{sec:discussion}

\subsection{Binary fraction}
The prediction of the number of \sbh binaries might be a lower limit due to the relatively low primordial binary fraction in the CMC models (5\%). For solar-type stars the initial binary/multiplicity fraction is likely higher \citep[30-40\%][]{2017ApJS..230...15M}. This results in a higher rate of binary-binary interactions providing an additional pathway to formation of \sbh in exchange interactions \citep[e.g.][]{Gonzalez2021, Tanikawa2024}. 

\subsection{Validity of the CMC catalogue}
The mass estimate of the progenitor cluster in the ED-2 stream, $\sim 5\times 10^4\units{\msun}$ \citep{Balbinot2024}, is lower than the least massive GC model in our catalogue ($1.2\times 10^5\units{\msun}$), so our results are based on extrapolation. However, they are within the same order of magnitude, so we expect that the values hold if we consider slightly lower cluster masses. Furthermore, the H\'enon-type CMC code is unable to resolve strict cluster dissolution, so the properties of the remaining binaries may change. Both points can be addressed with direct $N$-body models of the ED-2 progenitor.

\subsection{Stellar masses and detectability} 
The stellar masses in the overall population of \sbh in the CMC models peaks at relatively low masses ($\sim0.2\units{\msun}$), which makes the majority of these \sbh difficult to detect. Recent analyses of the stellar mass function Milky Way GCs \citep{2023MNRAS.521.3991B, 2023MNRAS.522.5320D} suggest that the low-mass end of the mass function could be flatter than  is assumed in CMC. This would lead to higher stellar masses in \sbh binaries. Estimating this effect requires additional dynamical modelling and is beyond the scope of this work \citep[see][for preliminary efforts]{Weatherford2021}. 

\subsection{Alternative pathways for \gb-like binary formation}
Recent work suggests that \gb-like objects may form via isolated binary evolution \citep{ElBadry2024, Iorio2024}. Although these models can accommodate the binary's intrinsic properties (including the lack of chemical pollution), they require a natal kick that is incompatible with \gb being associated to the ED-2 stream \citep{ElBadry2024} (but see \cite{Iorio2024} for alternative conclusions based on different stellar models). Although this argues for dynamical formation, the isolated channel is a promising additional pathway to \sbh formation. Our estimate of the dynamical production rates at $R_G\simeq8\,{\rm kpc}$ is only a factor of $\sim3-10$ higher than the isolated production rate of \citet{Iorio2024}, with the key difference being their radial distribution in the Galaxy. The dynamical formation scenario predicts that \sbh binaries follow the initial density distribution of GCs, $\propto R_G^{-4.5}$, whereas an isolated formation scenario predicts that these objects follow the shallower density profile of the stellar halo.

\section{Conclusions}
\label{sec:conclusions}
In this letter we studied the dynamical formation of \gb-like binaries in GCs. We analysed the public catalogue of star-by-star GC simulations in \cite{Kremer2020}, and compared it to \gb. We found that the properties of \gb\ (masses, period, eccentricity and lack of chemical pollution) are compatible with it being assembled dynamically from a previously unbound star and a BH that have not undergone joint stellar evolution. 

Furthermore, we extended our analysis to the whole ED-2 stream in particular and the Galactic halo in general. We predict that the stream produced $\sim 9 - 13$ dynamically assembled \sbh binaries, whereas the halo contains $\sim 10^5$ S-BH binaries, preferentially towards the Galactic centre. The next {\it Gaia} data release (DR4, late 2025) will allow us to test this prediction.

\begin{acknowledgements}
The authors thank Eduardo Balbinot for discussions on ED-2, Newlin Weatherford for useful feedback on the preprint, and the {\it Gaia} group at  ICCUB for useful interactions.
DMP and MG acknowledge financial support from the grants PID2021-125485NB-C22, EUR2020-112157, CEX2019-000918-M funded by MCIN/AEI/10.13039/501100011033 (State Agency for Research of the Spanish Ministry of Science and Innovation) and SGR-2021-01069 (AGAUR). SR acknowledges support from the Beatriu de Pin\'os postdoctoral program under the Ministry of Research and Universities of the Government of Catalonia (Grant Reference No. 2021 BP 00213). Support for KK was provided by NASA through the NASA Hubble Fellowship grant HST-HF2-51510 awarded by the Space Telescope Science Institute, which is operated by the Association of Universities for Research in Astronomy, Inc., for NASA, under contract NAS5-26555. 
\end{acknowledgements}

\bibliographystyle{aa} % style aa.bst
\bibliography{bibliography, bibliography_gaia} 

\end{document}